\documentclass[aps,prl,reprint,groupedaddress,longbibliography]{revtex4-1}
\usepackage{amsmath}

\usepackage{color}
\usepackage{braket}
\usepackage{bm}
\usepackage{graphicx}
\begin{document}

\title{Parallel low-loss measurement of multiple atomic qubits}

\author{Minho Kwon}
\email[]{mkwon22@wisc.edu}
\author{Matthew F. Ebert}
\author{Thad G. Walker}
\author{M. Saffman}
\affiliation{Department of Physics,
University of Wisconsin-Madison, 1150 University Avenue,  Madison, Wisconsin 53706}

\date{\today}

\begin{abstract}
We demonstrate low-loss measurement of the  hyperfine ground state of Rubidium atoms in a dipole trap array of five sites by state dependent fluorescence detection.
The presence of atoms and their internal states are minimally altered by utilizing circularly polarized probe light and a strictly controlled quantization axis.
We achieve mean state detection fidelity of \(97\%\) without correcting for imperfect state preparation or  background losses, and \(98.7\%\) when corrected. After state detection and correction for background losses, the probability of atom loss due to the state measurement is \(<2\%\)  and the initial hyperfine state is preserved with \(>98\%\) probability.
\end{abstract}

\maketitle

Experiments with qubits encoded in hyperfine states of neutral atoms are being actively developed as a route towards scalable quantum information processing\cite{Saffman2016}. Several different research groups have demonstrated preparation and control of order 50 qubits in 1D\cite{Endres2016}, 2D\cite{Xia2015,Barredo2016} and 3D\cite{YWang2016} optical lattices. Quantum computation requires qubit state measurements to determine the result of a computation, and for measurement based quantum error correction\cite{Devitt2013}. Measurement of the quantum state of an atomic hyperfine qubit is most often performed by using a cycling, or near cycling, transition which repetitively transfers the  qubit between a bright state $\ket{B}$ and an excited  state $\ket{e_B}$.
 Detection of scattered photons due to illumination with light that is near resonant with the cycling transition projects the qubit into state $\ket{B}$.
Conversely, if no photons are detected, the qubit is projected into the dark  state $\ket{D}$. This idealized picture breaks down if the cycling transition is not perfectly closed, in which case an atom in state $\ket{B}$ may suffer a Raman transition to $\ket{D}$ thereby giving a measurement error. 

Measurements that use a  cycling transition rely on the availability of  a  metastable qubit dark state $\ket{D}$,  or on shelving one of the qubit levels into a metastable dark state, as is done in trapped ion experiments\cite{Myerson2008}. 
In alkali atom experiments with qubits encoded in ground hyperfine levels 
the availability of a cycling transition generally relies on an angular momentum selection rule that is enforced by using probe light with a well defined polarization. This implies that the probe light propagates along a single axis in space which results in atomic heating due to the random direction of scattered photons. For a lossless measurement either the potential confining the atom should be sufficiently deep for the heating to be tolerable, as in experiments with trapped ions\cite{Haffner2008}, or the detection system should allow for a state measurement after scattering only a small number of photons to minimize heating. 
This latter approach was demonstrated with optically trapped
atomic qubits\cite{Fuhrmanek2011,Gibbons2011,Jau2016} using low noise single photon detectors. Alternatively, coupling of an atom to a high finesse cavity enables state detection with minimal heating and without loss of atoms\cite{Bochmann2010,Volz2011,HZhang2012}. It has  been proposed to perform fast state measurements by coupling a single atom to a many atom ensemble, as a means of increasing the effective photon scattering rate\cite{Saffman2005b}. It is also possible to enforce a dark state condition with three dimensional probing light that cools the atoms, but this requires an inconvenient and  complex sequence of steps\cite{Beterov2015}.

\begin{figure}[!t]
\includegraphics[width=8.5cm]{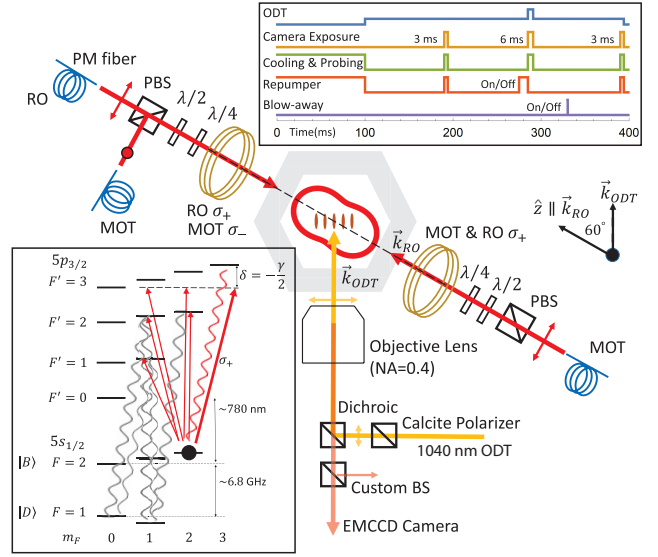}
 \caption{(color online)
  Experimental setup around hexagonal vacuum cell. The quantization axis $\hat{z} \parallel \vec{k}_{\rm RO}$ is set by the bias magnetic field from a pair of coils. $\sigma_+$ polarized light propagates along $\vec{k}_{\rm RO}$.
The horizontally polarized trapping light is in the plane formed by $\hat z$ and \(\vec{k}_{\rm ODT}\). Dichroic beamsplitters separate the trap light and fluorescence light which is imaged onto the camera.   }
  \label{fig_diagram}
\end{figure}

In order to take full advantage of the large number of qubits available in neutral atom experiments it is desirable to be able to losslessly measure multiple qubits in parallel. This can be done by imaging scattered light from an array of qubits onto a sensitive imaging detector such as an electron multiplying charge coupled device (EMCCD). Although EMCCD cameras have high quantum efficiency they suffer from excess readout noise which has hitherto rendered parallel lossless state detection infeasible. To circumvent this limitation previous array experiments used a ``blow away" technique where
atoms in $\ket{B}$ are ejected from the array using a single unbalanced beam, followed by detection of the presence or absence of an atom. Atom detection is performed using a 3D light field that cools the atoms, but does not prevent state changing Raman transitions during the measurement. This approach provides state measurements, but requires that a new atom will have to be reloaded, half the time on average, which severely impacts the experimental data rate.

In this letter we show that low-loss detection of multiple atoms, in parallel, is possible using an EMCCD camera. This  requires a careful choice of parameters to minimize both the motional heating rate (which is lower at large detuning) and the Raman depumping rate (which is lower at small detuning).     The enabling advances include use of a moderately high numerical aperture (NA) collection lens, deep optical traps, and careful preparation of the polarization state of the probe light to minimize Raman transitions
from $\ket{B}\rightarrow \ket{D}$. Similar results to ours have been independently reported in \cite{Martinez-Dorantes2017}.

\begin{figure}[!t]
    \includegraphics[width=8.5cm]{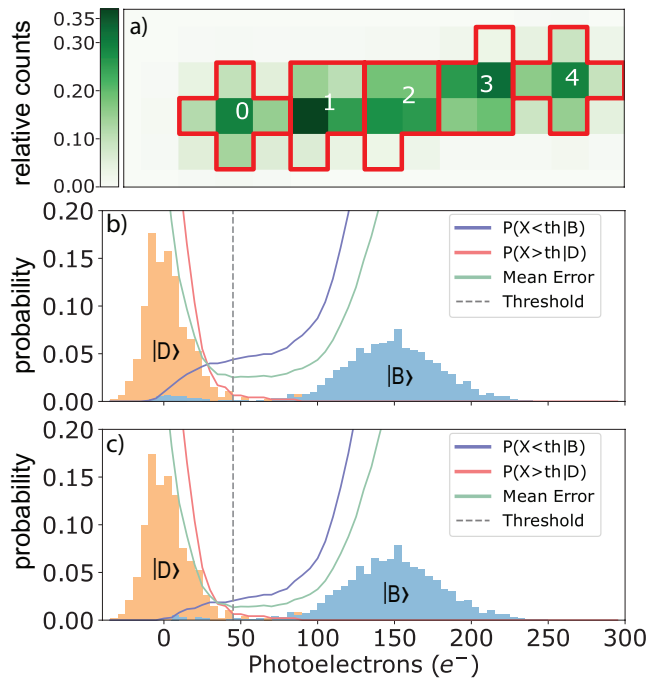}
  \caption{(color online)
  (a)   Regions of interest are five pixels enclosed by red borders with the relative photon counts on each pixel shown by the green shading. Each
5 pixel ROI receives \((76, 88, 89, 92, 76)\%\) of the light from the corresponding trapped atom. Neighboring site fluorescence crosstalk is \(\sim 2\%\). Each  pixel represents a \(4 ~\mu \rm m \times 4 ~\mu \rm m\) area and the site-to-site separation is \(\sim 9~ \mu \rm m\).
  (b) Histograms of non-destructive readout in the central region (\#2) for initial states $\ket{B}$ and $\ket{D}$ .
  (c) The same data set post-selected on the presence of an atom in the ROI in the third measurement, leaving only Raman depumping and state preparation as sources of error.
  Signals in histograms are background-subtracted.
  }
  \label{fig_hists}
\end{figure}

The experimental geometry and measurement sequence are  shown in Fig. \ref{fig_diagram}.
Atoms are prepared  in the $\ket{F=1}$ or $\ket{F=2}$ hyperfine  levels of the  $^{87}$Rb $5s_{1/2}$ electronic ground state,
corresponding to $\ket{D}$ and $\ket{B}$ respectively.
To prepare states of single atoms we begin by cooling  in a standard magneto-optical trap (MOT) that is then overlapped with a 1D array of five optical
dipole traps (ODTs) formed by focusing 1040 nm light to a waist of
$w\simeq 2.5~\mu\rm m$. The traps are (2.8, 4.4, 5.6, 3.9, 3.4) mK  deep and are spaced by $\sim 9~ \mu\rm m$.  The traps are pencil shaped with sizes  $\sigma_{z}\sim 7 \mu {\rm m}, \sigma_{r}\sim 0.7\mu \rm m$, with the long axis along the optical axis of the collection optics. Single atoms are loaded with probability 20-30\% at a temperature of $\sim 100~\mu\rm K$.

In order to measure the initial trap populations, the atoms are probed
 using 6 MOT beams with components near-resonant with \(\Ket{B}\leftrightarrow\Ket{e_B}\) and \(\Ket{D}\leftrightarrow\Ket{e_D}\) simultaneously, where
$\ket{e_B}$ is the $F'=3$ level and $\ket{e_D}$ is the $F'=2$ level of the $5p_{3/2}$ excited state.
Atom fluorescence is collected by a $NA=0.4$ lens, and imaged onto an EMCCD camera (Andor iXon EM+ DU-860).
The magnification was chosen such that the site separation is 2 pixels, and the signal from each ODT is integrated over a region of interest (ROI) defined by 5 camera pixels, as shown in Fig. \ref{fig_hists}a).
The advantage of a low magnification is that the same signal can be integrated over fewer pixels which lowers the electronic background noise.
The excited states, $\ket{e_B},\ket{e_D}$, are anti-trapped in the ODT, so to avoid heating the atom we toggle the ODT and the probe beams out of phase with a 50\% duty cycle at 1.25 MHz. The photon detection efficiency is estimated to be  \(1.6-2.0\%\),
accounting for the lens solid angle and dipole emission pattern (3.9 \%),
transmission through optics (74\%), EMCCD quantum efficiency ($\eta = 75\%$), and fluorescence lying outside of the camera pixels used to define regions of interest (76-92\%).

Upon completion of the population measurement, there is a 100 ms delay for image transfer to the computer, after which the atoms are initialized in a random superposition of the Zeeman substates of one of the hyperfine levels, chosen by leaving either \(\Ket{D}\leftrightarrow\Ket{e_D}\) or \(\Ket{B}\leftrightarrow\Ket{e_B}\) on to depopulate the coupled state.
To prevent low intensity leakage light from disrupting the initialized states mechanical shutters block unwanted light after initialization is completed.
We estimate the state preparation fidelity for both states to be \(> 99.5\%\)
limited by the fidelity of blow away measurements that are performed at reduced ODT depth.

After state initialization, a bias magnetic field \(B_{z}\sim 20 ~\text{G}\) making an angle of 60\({}^{\circ}\) from $\vec k_{\rm ODT}$, the long axis of the ODTs, is switched on. The beams used for probing propagate along and counter to  $\vec k_{\rm RO}$,  which is  set to be parallel to $\hat z$ with a possible small alignment error $\theta$, see Fig. \ref{fig_diagram}.
We use counter-propagating probe beams to mitigate the effect of heating due to near-resonant radiation pressure.
In order to suppress Raman transitions both readout beams are $\sigma_{+}$ polarized  which optically pumps the
atoms into the  lower state  of the $\ket{2,2}\leftrightarrow\ket{3',3'}$ cycling transition.
The beams are circularly polarized with small measured impurity of \(\sim 6.3\times 10^{-4}\) \cite{Kwon2017SM}.
The counter-propagating probe beams are generated from separate lasers with a relative frequency offset of 500 kHz.
This technique avoids standing wave patterns, which can cause a time dependent drift in the single atom scattering rate and broaden the single-atom camera signal distribution.
During the state measurement sequence the trap depths are temporarily doubled  to enhance retention of the atoms.
The combined intensity and detuning of the probe beams is set to saturation parameter \(s_{0}=1\)  and \(\delta = \frac{\gamma}{2}\) red of the Zeeman shifted  $\Ket{2,2}\leftrightarrow\Ket{3',3'}$ transition to provide maximal damping\cite{Wineland1979} with $\gamma$ the excited state linewidth.
The atoms are illuminated for 6 ms with the same 50\% duty cycle as is used for the population measurement and  fluorescence light is collected by the EMCCD for analysis.
The resulting data are shown in Fig. \ref{fig_hists}. The hyperfine state is determined on the basis of a simple threshold condition relative to the vertical dashed lines in Fig. \ref{fig_hists}b),c). Although more extensive analysis that utilizes information gained from the temporal or spatial distribution of light in each region of interest can further reduce uncertainties\cite{Myerson2008,Martinez-Dorantes2017} our results show that the threshold condition alone is adequate for high fidelity measurements.

After an additional  100 ms delay for image transfer,  a third readout sequence probes the atoms again.
Depending on the experiment, the third readout is either a second
population measurement for probing atom loss or a destructive state selective measurement  using a blow away beam for measuring the number of atoms depumped from \(\Ket{B}\) to \(\Ket{D}\).
Full characterization of the non destructive measurement requires 4 experiments: 2 (state preparation  \(\Ket{B}\) or \(\Ket{D}\)) \(\times\) 2 (blow away on or off).
The results of the 4 experiments for the center site are summarized in Table \ref{table_Fidelity} for the center site and Table \ref{tab.2} for the other sites. 
We note that the results marked with a) include  2 \% atom loss between each camera readout due to the finite trap lifetime \(\tau \sim 5\) s and the 100 ms gap between each measurement.
The background collision loss is not a fundamental limitation, and could be reduced  by decreasing the chamber pressure or by shortening the image transfer time.

\begin{table}[!t]
\centering
\caption{Results in the central site  averaged over   2000 measurements.
Data marked (a) are without correction, and data marked (b) are post-selected  on the presence of an atom in the ROI in the third measurement, leaving only Raman depumping and state preparation as sources of error. The final state results are found from a third, state-selective measurement using a blow away beam.
}
\label{table_Fidelity}
\begin{tabular}{|c|c|c|c|c|c|}
\hline
 & \multicolumn{2}{c|}{detected state (\(\%\))}  &  \multicolumn{3}{c|}{final state (\(\%\))} \\
 \hline
initial state    & \(\ket{B}\)  & \(\ket{D}\)   & \(\ket{B}\)   & \(\ket{D}\) & Lost   \\
\hline
\(\ket{B}\) & \begin{tabular}[c]{@{}l@{}} (a) 95.6(6)\\(b) 98.0(4)\end{tabular}
& \begin{tabular}[c]{@{}l@{}}(a) 4.4(6)\\(b) 2.0(4)\end{tabular}
& 98.6(1.9) & 0.6 (1.6) & 0.8(1.3)\\
 \hline
\(\ket{D}\) & 0.6(4)   & 99.4(4) & N/A & 99.6(1.6) & 0.4 (1.6)  \\
 \hline
\end{tabular}
\end{table}

\begin{table}[!t]
\centering
\caption{Loss-corrected detection fidelities for the other four shallower traps. \(\ket{\psi}_{i}\) is the initially prepared state.}
\label{table_Fidelity_4sites}
\begin{tabular}{|c|c|c|c|c|c|c|c|c|}
\hline
			& \multicolumn{8}{c|}{Detected states (\%)}                                                                      \\ \hline
ROI           & \multicolumn{2}{c|}{\#0} & \multicolumn{2}{c|}{\#1} & \multicolumn{2}{c|}{\#3} & \multicolumn{2}{c|}{\#4} \\ \hline
\(\ket{\psi}_{i}\)            & \(\ket{B}\) &\(\ket{D}\) & \(\ket{B}\) &\(\ket{D}\) & \(\ket{B}\) &\(\ket{D}\) &\(\ket{B}\)  &\(\ket{D}\)  \\ \hline
\(\ket{B}\)   &   97.1(5)    &  2.9(5)       &    98.3(3)   &   1.7(3)  &    97.7(6)  &   2.3(6)  &   98.2(1.2) & 1.8(1.2)  \\ \hline
\(\ket{D}\)    &    0(0)     &   100(0)      &    1.0(5)    &   99.0(5) &     0.5(4)  &  99.5(4)  &      0      &  100(0)   \\ \hline
\end{tabular}
\label{tab.2}
\end{table}

The primary limitation to the non-destructive measurement  is the mean number of photons, $N_\gamma$, that can be scattered before the atom is depumped from $\Ket{B}$ to $\Ket{D}$.
When using random polarization  \(N_{\gamma}=\frac{38340}{1+4\delta^2/\gamma^2+s_{0}}\), where \(s_{0}=I/I_{\rm s,eff}\) and  \(I_{\rm s,eff}=3.6 ~ \textrm{mW}/\textrm{cm}^2\) is the saturation parameter for randomly polarized light, see  \cite{Kwon2017SM} for a derivation.
With typical experimental parameters $10^4$ photons could be scattered which would lead to approximately 100 photo-electrons, which is technically enough to clearly resolve the $\Ket{B}$ and $\Ket{D}$ photon histograms.
However the $\Ket{B}$ state histogram would leave a long tail from depumping events during the exposure that would overlap with the
$\ket{D}$ state  distribution.
Therefore, in order to obtain clearly distinguishable photo-electron statistics we need the additional constraint that atoms scatter \(\sim 10^4\) photons with minimal depumping, a condition that isotropic polarization does not satisfy.

\begin{figure}[!t]
   \includegraphics[width=8.4cm]{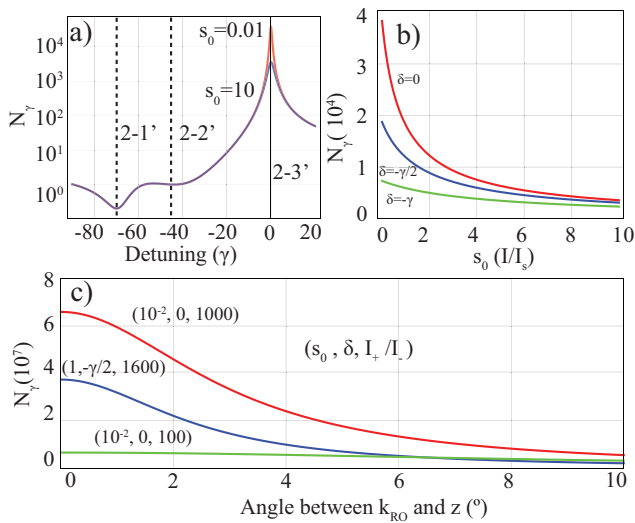}
  \caption{(color online) Dependence of mean number of resonant photons per Raman photon on probe light parameters.
  a) Detuning dependence spanning neighboring levels.
  b) Intensity dependence at three different detunings.
  c) Enhancement with $\sigma$ polarized light
for given saturation, detuning and intensity contrast $I_+/I_-$
between $\sigma_+$ and $\sigma_-$.
}
  \label{fig_depumping}
\end{figure}

To suppress the depumping we have used  $\sigma_+$-polarized light along the quantization axis, as described above.
In a real experiment polarization impurities and a small angular mismatch $\theta$ between $\hat z$, the direction of the magnetic field,
and $\vec k_{\rm RO}$, the axis of the readout beams, will still allow for a finite depumping rate.
The figure of merit is the number of photons that the bright state can scatter before it falls into the dark state, as shown in Fig. \ref{fig_depumping}.
We can quantify the chance of depumping by summing the rates over Raman depumping channels and comparing  to the scattering rate on the  cycling transition\cite{Kwon2017SM}.
We estimate that we are able to scatter \(N_{\gamma,\sigma}=3.7\times 10^{5}\) photons corresponding to an enhancement factor of \(\sim 20\) over the unpolarized case  with parameters $s_{0}=$1, \(\delta=-\gamma/2\), and measured polarization purity $I_{\sigma_+}/I_{\sigma_-}=1600$.

It is also necessary to consider depumping due to the time-dependent vector and tensor light-shifts imposed by the ODT.
Circular polarization of the ODT light results in  a vector shift on the atoms which adds a fictitious magnetic field, \(\vec{B}_{\rm fict}\), along \(\vec{k}_{\rm ODT}\), which can be expressed as \(\vec{B}_{\rm fict}=\frac{U_0}{\mu_{B}g_{\ket{B}}}\left[\frac{\mathcal{A}\alpha^{(1)}_{\ket{B}}}{\alpha^{(0)}_{\ket{B}}}\right]\hat{k}_{\rm ODT}\), where $U_0$ is the trap depth, \(g_{\ket{B}}\) is the Land\'e g-factor, and \(\alpha^{(0)}_{\ket{B}}\), \(\alpha^{(1)}_{\ket{B}}\) are scalar and vector polarizabilities.
Circularity of polarization is characterized by \(\mathcal{A}\) which ranges between \(-1\leq\mathcal{A}\leq 1\).
The \(60^{\circ}\) angle between $\vec{k}_{\rm ODT}$ and $\vec{k}_{\rm RO}$ means that Larmor precession occurs, which reopens the depumping channels.
In terms of the trap depth $U_0$ the fictitious field is $\vec{B}_{\rm fict}/U_0 = 29.77 \left[\frac{\mathcal{A}\alpha^{(1)}_{\ket{B}}}{\alpha^{(0)}_{\ket{B}}}\right]\hat{k}_{\rm ODT}$ G/mK {\cite{Kwon2017SM}} .
For our experimental parameters, \(\mathcal{A}\sim 2\times 10^{-4}\), \(\lambda=1040 ~\text{nm}\), \(\vec{B}_{\rm fict}/U_0=0.3\quad\text{mG}/\text{mK}\).
Therefore if one works with several mK deep traps and a weak bias field, a nonzero angle between $\vec k_{\rm RO}$ and $\vec k_{\rm ODT}$   must be accounted for.
We verified that the depumping rate was independent of ODT power\cite{Kwon2017SM}, suggesting that depumping is not due to the vector light shift for our parameters.

In addition, tensor light shifts couple  \(M_{F}\) states in the excited level, creating a new set of energy eigenstates that are superpositions of \(\ket{F,M_{F}}\) states, which breaks the cycling character of the $\ket{2,2}\leftrightarrow \ket{3',3'}$ transition.
To avoid tensor shifts during readout the probe laser and ODT are chopped out of phase so that the excited state is never populated when the ODT is on.

\begin{figure}[!t]
   \includegraphics[width=8.5cm]{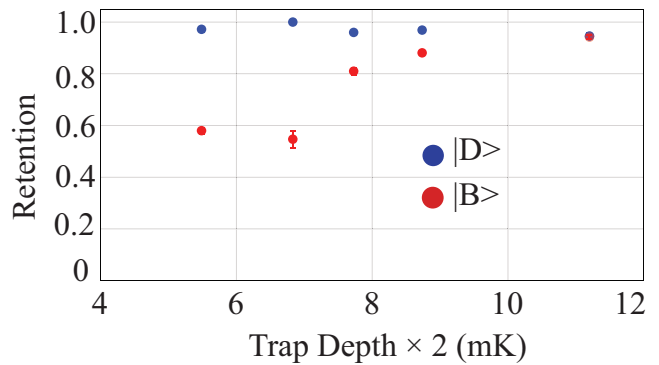}
  \caption{(color online)
  Probability of atom retention  after non-destructive readout as a function of trap depth.
  Background gas collisions cause \(\sim 4\%\) atom loss between the first and third measurements.
}

  \label{fig_postSSRO}
\end{figure}

In the experiments reported here the  Zeeman states in each hyperfine level are equally populated, therefore depumping can occur in transit while optically pumping to the stretched state.
This transient depumping is present even with perfect polarization and alignment. On average the amount of depumping will be very similar to what is expected for atoms prepared in  $M=0$  states, so  our results are representative of measurements of qubit states. The relatively strong field \(B_{z}\sim 20\) G, used to suppress effects from the vector light shift, causes the $m_F$ states to have non-optimal detunings for transitions \(\Ket{2,M_{F}}\rightarrow \Ket{3', M_{F}+1}\) for $M_F\ne 2$ due to the Zeeman shifts.
Decreasing the bias field will minimize the transient state depumping, but
we expect this to add to our state detection error at the \(0.2-0.5\%\) level.
This can be estimated by $n_{op}/N_{\gamma}$ where \(n_{op} \sim 10\) is the mean number of scattered photons to optically pump into $\Ket{2,2}$ and \(N_{\gamma}\) is a few thousand when not in the cycling state \cite{Kwon2017SM}.

Despite the use of counter-propagating $\sigma_+$ beams, heating was still noticeable, limiting atom retention after the measurement, as is shown in Fig. \ref{fig_postSSRO}, and forcing us to use traps that are $\sim10$  mK deep. This limited performance may be attributed to laser intensity noise, lack of sub-Doppler cooling mechanisms, and 1-D cooling. Future improvements including working with a higher NA lens to reduce the number of scattered photons needed for a measurement, and cooling the atoms into the Lamb-Dicke regime to suppress recoil heating will further reduce atom loss. Using blue detuned traps with intensity minima at the location of the atoms, as in \cite{Piotrowicz2013,YWang2016}, would  reduce the excited state tensor mixings, and obviate the need to turn the ODT on and off, thereby reducing any heating due to trap switching.

Ideally, qubit measurements should be projective, leaving the atom in an eigenstate of $\sigma_z$. This can be accomplished by following  detection
 of an atom in $\ket{B}$,  which ends up in $\ket{2,2}$, with a sequence of stimulated Raman transitions or microwave pulses to return the atom to  the $\ket{2,0}$ hyperfine state.

In conclusion we have demonstrated non-destructive parallel readout of an array of five Rb atoms. Increasing the collection efficiency of the imaging optics, combined with colder atoms, and possibly more refined analysis of the spatial information provided by the camera, we anticipate that loss of atoms due to heating can be reduced to a level compatible with implementation of repetitive error correction for quantum computation.

This work was supported by NSF award 1521374, the AFOSR Quantum memories MURI,
 and  the ARL-CDQI through through cooperative
agreement W911NF-15-2-0061. MS thanks Dieter Meschede for sharing their results prior to publication.


%


\newpage

\pagebreak

{\center\bf Supplementary Material for \\ Parallel low-loss measurement of multiple atomic qubits}

\section{Fluorescence statistics with atom  loss}
\begin{figure}[!t]
  \includegraphics[width=240pt]{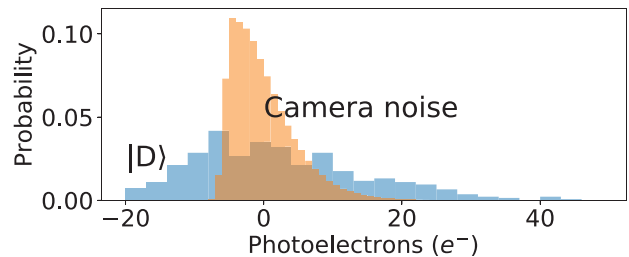}
  \caption{
    A comparison of the intrinsic camera noise in the 5 pixel region of interest (ROI) with the camera shutter closed (yellow), and the photo-electron distribution for $\Ket{D}$ (blue).
    The data are centered about the mean of the distribution. Intrinsic camera noise is \(\sim 22 (e^{-})^{2}\), while the dark state has variance of \(\sim 196 (e^{-})^{2}\), due to additional noise in the experimental setup.
  }
  \label{fig_roibg}
\end{figure}

With no loss mechanisms, the camera signal distributions for the cases of bright $\Ket{B}$ and dark $\Ket{D}$ states after probing for a time $t$ are given by Poissonian distributions with means
\begin{equation}
	\begin{split}
    	\mu_D(t) &= (\gamma_D + \gamma_{bg})t+\mu_{CIC}, \\
    	\mu_{B}(t) &= (\gamma_B +\gamma_{bg})t+\mu_{CIC},
    \end{split}
\end{equation}
where $\gamma_{bg}$, $\gamma_D$, and $\gamma_B$ are the background, dark state, and bright state photo-electron production rates and $\mu_{CIC}$ is the background photo-electron rate due to clock induced charge (CIC).
CIC is a Poissonian noise source intrinsic to EMCCD cameras and is independent of the exposure time.
It is discussed in more detail in the next section.
The photo-electron production rate from $\Ket{D}$, given by $\gamma_D$, is negligible compared to $\gamma_B$ and $\gamma_{bg}$, therefore we set $\gamma_D=0$ for this section and consider the dark state distribution as a background distribution for the bright state.
Both the large average number of photo-electrons, $\mu_D \sim 100$, and fluctuations in probe intensity and detuning broaden the expected single-atom signal.
Therefore we can treat the photo-electron distributions, $S_{B}(s)(S_D(s))$, as Gaussian: $G(s,\mu,\sigma) = (2\pi\sigma^2)^{-1/2}e^{-(s-\mu)^2/2\sigma^2}$.
For our system, the effect of spurious noise from CIC is lower than other sources of background $\gamma_{bg}t>\mu_{CIC}$ for our exposure times, therefore we can simplify the analysis by assuming a Gaussian distribution for all noise sources, see Fig~\ref{fig_roibg}.
The width of the distributions, $\sigma_B(\sigma_D)$, are determined experimentally by fitting the distributions given by
\begin{equation}
	\begin{split}
		S_D(s) &= G(s,\mu_D,\sigma_D), \\
        S_B(s) &= G(s,\mu_B, \sqrt{\sigma_B^2+\sigma_D^2}),
	\end{split}
\end{equation}
to the relevant camera signal distributions with no loss.
The width of the background must be deconvolved from the width of the bright state distribution to correctly include the effect of losses during the measurement, although typically $\sqrt{\sigma_B^2+\sigma_D^2} \approx \sigma_B$.
For $\Ket{B}$ a lossless measurement can be done by leaving the $\Ket{D}\leftrightarrow \Ket{e_D}$ hyperfine repumping light on during the camera exposure.

When losses during readout are included the bright state distribution, $S_B(s)$, changes from Gaussian to something more complicated.
If the atom in $\Ket{B}$ is lost or depumped into $\Ket{D}$ at time $t'<t$, then the atom will cease scattering photons and will only accumulate photo-electrons at $\gamma_{bg}$ for a time $t-t'$.
Therefore the mean signal for an atom initially in $\Ket{B}$ undergoing a loss event at time $t'$ is given by $\mu_B^{\star}(t') = \gamma_B t' + \gamma_{bg} t + \mu_{CIC}$.
This effectively adds a tail to the ideal Gaussian distribution.
The normalized tail distribution, $S_B^\star$, is given by
\begin{equation}
	\label{eqn_sbstar}
		S_B^{\star} = \left(\frac{\alpha}{1-e^{\alpha t}}\right)\int_0^t dt'e^{-\alpha t'}G(s, \mu_B^{\star}(t'), \sigma_B^\star(t')),
\end{equation}
where $\alpha$ is the combined heating induced loss and depumping rate and $\sigma_B^\star(t') \equiv \sqrt{\sigma_B^2(t'/t)+\sigma_{D}^2}$.
To the best of our knowledge, this integral cannot be solved analytically
unless $\sigma_{bg}=0$.
For simplicity we also set $\gamma_{bg},\mu_{CIC}=0$ and Eq. (\ref{eqn_sbstar}) becomes
\begin{equation}
	\begin{split}
	S_B^\star &\simeq \frac{\alpha}{2(1-e^{-\alpha t})}\sqrt{\frac{t}{\chi}}e^{s\gamma_B t/\sigma_B^2}\left[A_+ - A_-\right],
    \end{split}
\end{equation}
where $\chi\equiv \gamma_B^2t+2\alpha\sigma_B^2 $ and
$$A_\pm \equiv e^{\pm (s/\sigma_B^2)\sqrt{\chi t}}\left[\mathrm{erf}\left(\frac{s\pm\sqrt{\chi t}}{\sqrt{2}\sigma_B{}}\right)-1\right].$$
The original distribution with finite background can be recovered by convolving $S_B^\star$ with the background distribution $G(s,\gamma_{bg}t+\mu_{CIC},\sigma_{D})$.
In the case of small loss this is a small effect and the time-intensive convolution operation is not necessary.

The full camera signal distribution model, $S_B$, can be obtained now by a weighted sum of the distribution with no loss, $S_B^{(0)}\equiv G(s,\mu_B,\sigma_B)$, and the tail distribution with a loss event, $S_B^\star$:
\begin{equation}
  \label{eq_lossmodel}
  S_B(s) = e^{-\alpha t}S_B^{(0)}(s) + (1-e^{-\alpha t})S_B^\star(s)
\end{equation}

\begin{figure}
  \includegraphics[width=240pt]{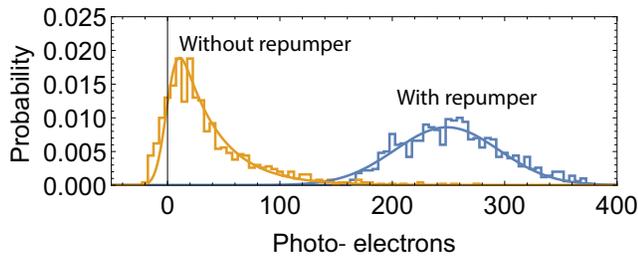}
  \caption{
    Signal from a $t=4$ ms exposure of $\Ket{B}$ with isotropic polarization, detuning $\delta \sim -1.5 \gamma$, and intensity $s_{0} \sim 10$, with (blue) and without (yellow) the extra $\Ket{D}\leftrightarrow\Ket{e_D}$ hyperfine repumping light.
    The blue solid line is a fit to a Gaussian distribution with no loss, and is used to extract parameters for the fit to the yellow curve.
    The yellow curve is a fit for the loss rate $\alpha$ in Eq. (\ref{eq_lossmodel}) convolved with the background Gaussian distribution, all other parameters are fixed.
    The result of the fit yields a depumping rate of $\alpha=1.8(1)$ ms$^{-1}$.
    The large loss rate for the case of isotropic polarization emphasizes the necessity of strict polarization control.
  }
  \label{CameraSignal_Example}
\end{figure}

An example histogram for $\Ket{B}$ is shown under conditions of large depumping losses (yellow) compared to no depumping loss (blue) in Fig~\ref{CameraSignal_Example}.

\section{Camera noise}
EMCCD cameras have multiple sources of noise which broaden the camera count distribution including dark counts, CIC, EM gain register noise, and analog to digital converter (ADC) noise.
Modeling is possible though complicated\cite{Hirsch2013}.
Cooling the detector to -70 C reduces dark counts to a rate that is negligible on the scale of our $t<10$ ms camera exposures.
CIC is a Poissonian process, caused by impact ionization when reading out camera pixels, that gives the background photo-electron signal a long tail, see Fig~\ref{fig_roibg}.
The long CIC tail is independent of exposure time and therefore sets a limit on how many photo-electrons must be collected to make a high fidelity threshold-based measurement.
CIC events should lack any spatial correlation with photo-electron events, so the effect could be reduced further by including spatial information, such as an auto-correlation, into a multi-dimensional threshold cut. 

By taking images with the shutter closed, the unavoidable CIC and ADC noise level of the camera  is found to have a standard deviation of $\sim 5$ photo-electrons.
The effect of environmental noise sources such as room lights, probe scattering from surfaces, and fluorescence from untrapped background atoms are minimized by a narrow bandpass interference filter and a spatial filter, but still contribute to the background at a combined rate of 29 photo-electrons/ms averaged over the exposure time.
Arbitrarily selecting more pixels for ROI will add these noise sources on top of the signal and broaden the histogram peaks, limiting the maximal fidelity.
Fluorescence from neighboring sites can contribute to the signal as well. This crosstalk is \(2\%\) of the single atom signal per site. For instance with a \(\sim 150~ e^{-}\) single atom signal, the neighboring site shows a  \(\sim 3 ~e^{-}\)  signal.

\section{Determination of region of interests}

\begin{figure}[!t]
\includegraphics[width=200pt]{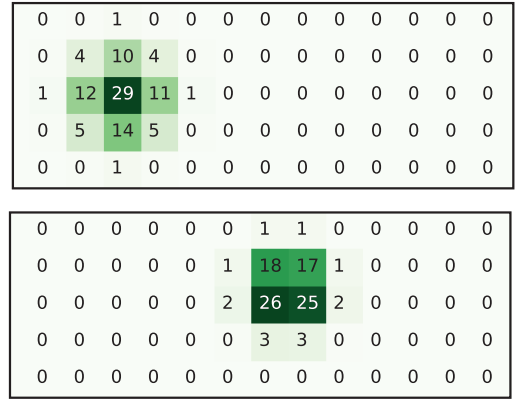}
\caption{Fluorescence from individual atoms resolved by ICA. Numbers are the percentage of normalized signal received by that pixel. ICA results for ROI\#0(upper) and \#2(lower) are shown.}
\label{fig_ICA}
\end{figure}

In a noise-free detector all pixels that contain signal could be included in the ROI.
In practice detector noise  prevents this because including more pixels  leads to more noise.
To maximize Signal-to-Noise Ratio (SNR), a few dominant pixels are chosen for the ROI.
Since our trap location is stable, we can set the regions of interests based on high SNR optical molasses imaging.
We take several thousand camera shots containing stochastic loading events on five sites.
Since atom loading events are uncorrelated we perform independent component analysis (ICA)\cite{Hyvarinen2000} to infer the locations of the independent emitters.
The result of the analysis is shown in Fig. \ref{fig_ICA}.
Most of the signal from each site is  localized onto 4-5 pixels.

Intrinsic camera noise is uniform over pixels so the number of included pixels determines the noise contribution from the camera. To keep the effect the same at all  five sites we choose the same number of pixels at each site. The maximum number of pixels that gave us non-overlapping ROIs was five, which contained \(76-92 \%\) of the photo-electron counts  based on the ICA.

\section{Dipole emission pattern}
The dipole emission pattern is not isotropic, and therefore a simple solid angle estimate based on the lens numerical aperture (NA) is not sufficient.
When emitting a circularly polarized photon, the emission pattern is that of a rotating dipole.
The fraction of light collected by a lens with numerical aperture NA is:
\begin{equation}
  \text{CE}=\int_{\theta_{i}}^{\theta_{f}}\int_{-\phi_{0}}^{\phi_{0}}\frac{3}{16\pi}\left(\cos^{2}{\theta}+1\right)d\phi \sin{\theta}d\theta
\end{equation}
where CE denotes collection efficiency, \(\theta\) is polar angle from the atom's quantization axis and \(\phi\) is azimuthal angle.
For a lens with a given NA, \(\theta_{i}=\frac{\pi}{2}-\arcsin{(\text{NA})}\), \(\theta_{f}=\frac{\pi}{2}+\arcsin{(\text{NA})}\), \(\phi_{0}=\arcsin{(\sqrt{\frac{\text{NA}^2}{\sin^2{\theta}}-\frac{1}{\tan^{2}{\theta}}})}\).
In our configuration, the quantization axis makes an angle $\alpha$ with the optical axis of collection.
Therefore in the lens' spherical coordinates the integrand in parentheses becomes
\begin{equation}
  \cos^{2}{\theta}\rightarrow \left[\cos{(\alpha)}\cos{(\theta)}-\sin{(\alpha)}\sin{(\theta)}\cos{(\phi)}\right]^{2}
\end{equation}
In our setup \(\alpha=60^{\circ}\) and \(\text{NA}=0.40\), which yields a collection efficiency of \(\text{CE}=3.94\%\) which is slightly less than for the  case of isotropic emission which would give  \(\text{CE}=4.17\%\).

\section{Optimization of Light Polarization}

Standard dielectric cube polarizing beam splitters (PBSs) are used to set a linear polarization, and a pair of \(\lambda/4\), and \(\lambda/2\) retarders map the linear polarization to circular. Although in principle a \(\lambda/4\) retarder is sufficient to map linear polarization to circular, we found that the use of an extra \(\lambda/2\) retarder provided better adjustability leading to higher polarization quality. The quality is characterized by a rotating polarizer followed by a photodetector after the beam passes through the vacuum cell.
Circularly polarized light transmits regardless of the PBS orientation, while linear polarization does not.
Qualitatively the more circular the light is, the smaller the oscillation amplitude as the detection polarizer is rotated.
The retarders are rotated to minimize the  amplitude of the oscillation.
The contrast ratio of the visibility for maximally linear and maximally circular polarizations is used to quantify the purity of the polarization.
Assume the polarization state has equal magnitude Cartesian components \(E_{x}=E_{y}=E_{0}\) with a finite phase difference \(\phi\), represented as \(\bm{E}=\begin{bmatrix}
E_{0},&E_{0}e^{i\phi}
\end{bmatrix}\) the intensity after passing through the rotating PBS is
\begin{equation}
I(\phi,\theta)/I_{0}=\frac{1}{2}\left[1+\cos(\phi)\sin(\theta)\right]
\end{equation}
where \(\theta\) represents the rotation angle of the PBS. The amplitude of the modulation is \(\cos(\phi)/2\) which determines the relative phase.
We now decompose the original electric field \(\bm{E}\) into $\sigma_+$ and $\sigma_-$   as \(E_{\sigma_+}=\begin{bmatrix}
\frac{1}{\sqrt{2}},&-\frac{i}{\sqrt{2}}
\end{bmatrix}\cdot\bm{E}\),  \(E_{\sigma_{-}}=\begin{bmatrix}
\frac{1}{\sqrt{2}},&\frac{i}{\sqrt{2}}
\end{bmatrix}\cdot\bm{E}\).
Therefore the intensity ratio or polarization purity is
\begin{equation}
\frac{I_{\sigma_+}}{I_{\sigma_-}}=\left|\frac{E_{\sigma_+}}{E_{\sigma_-}}\right|^{2}=\frac{1+\sin(\phi)}{1-\sin(\phi)}
\end{equation}
With the rotating PBS setup, the DC and AC value of the intensity variation can be easily measured. We define the contrast as \(C=\frac{\text{DC}}{\text{AC amplitude}}=\frac{1/2}{\cos({\phi}/2)}\).
Then the resulting intensity ratio is
\begin{equation}
\frac{I_{\sigma_+}}{I_{\sigma_-}}=\frac{1+\sqrt{1-1/C^{2}}}{1-\sqrt{1-1/C^{2}}}
\end{equation}
and can be approximated as \(4C^{2}\) for \(C\gg1\).
Using generic polarizing optics we achieve contrasts of $C=10-20$, corresponding to polarization purities of 400-1600. Higher grade optics can readily improve the extinction. For example we have observed \(C\sim50\) with Glan-Taylor polarizers corresponding to \(10^{4}\) polarization purity.

\section{Magnetic field optimization}
\begin{figure}
  \centering
  \includegraphics[width=225pt]{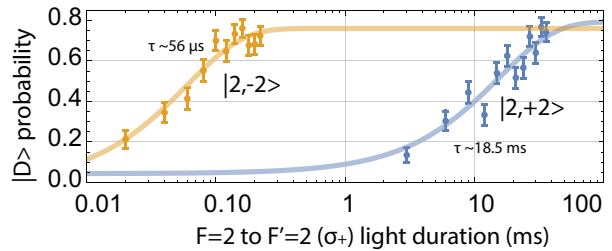}
  \caption{
  	Growth of the probability of being depumped to the dark state \((F=1\)) for two different stretched Zeeman states \(\Ket{2,+2}\) (blue) and\(\Ket{2,-2}\) (yellow).
    The incident probe is unidirectional \(\sigma_{+}\), tuned to the center of the transition \(\ket{2,0}\rightarrow \ket{2',0'}\) under bias field \(\sim 5\) G.
  }
  \label{fig_depumping_SM}
\end{figure}

The magnetic field must be closely matched to the probe light polarization.
We adapt a procedure from reference \cite{Martinez2016thesis}, which uses the atoms to optimize the magnetic field vector, which defines the quantization axis $\hat{z}$, to coincide with  $\vec{k}_{\rm RO}$ defined by the propagation vector of the probe light.
The atoms are first optically pumped to \(\Ket{F=2,m_{F}=2}\) by a weak, circularly polarized, unidirectional beam with \(B_{z}\sim 5~ \text{G}\).
{One of the $\sigma_+$-polarized probe lasers} is tuned to \(\Ket{2}\leftrightarrow\Ket{2'}\) and optically pumps the atoms into the $\Ket{2,2}$ dark state.
When the alignment is optimal, the $\Ket{2,2}$ state is nearly dark and can only couple off-resonantly to \(\Ket{3',3'}\), so depumping to $\Ket{1}$ is minimized.
If there is any mismatch, the dark state mixes with the bright states and scatters photons, eventually depumping into \(\Ket{1}\), which can be measured by the destructive blow away measurement.
The growth in time of \(\Ket{D}=\Ket{1}\) as a function of the depumping light \(\Ket{2}\rightarrow \Ket{2'}\) quantifies the quality of the alignment, see Fig. \ref{fig_depumping_SM}.
The depumping time constants can be compared by preparing \(\Ket{2,-2}\) using the other MOT $\sigma_-$ beam, and repeating the sequence.
The ratio between the time constants can be used as a figure of merit for the alignment. We measure a ratio \(\sim 330 \) as seen from Fig. \ref{fig_depumping_SM}

\section{Transient depumping}

\begin{figure}
  \includegraphics[width=225pt]{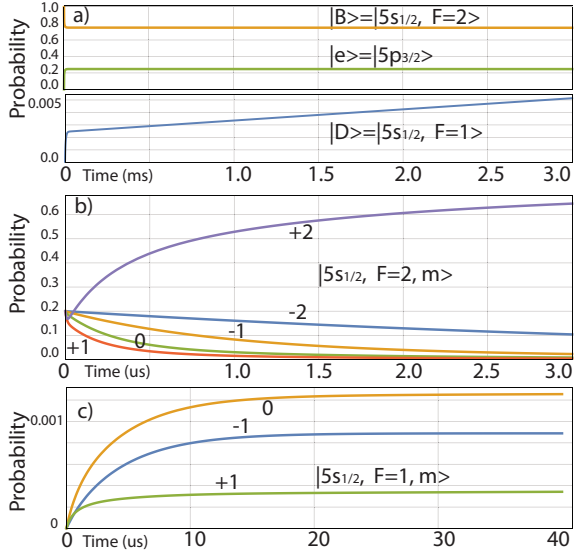}
  \caption{State dynamics during readout. 	
  a) Probability of being in ground hyperfine levels and the excited level. Dark state probability is shown on a  different y-scale.
  b) Transient optical pumping dynamics for the bright state manifold during readout. An initially random $\Ket{2,m_F}$ state population is  pumped to the stretched state. 
  c) Transient dark state dynamics. Off-resonant coupling to $\Ket{2'}$ is possible for all $m_F$ states $m_F \neq2$, so an increased Raman rate is expected until the atom is in $\Ket{2,2}$.
    }
  \label{fig_transient}
\end{figure}

When starting from the non-stretched state a higher depumping rate is expected until the atom has been pumped into $\Ket{2,2}$, since the transitions can off-resonantly couple to $\Ket{2'}$.
A numerical simulation to estimate this source of error has been performed using our experimental parameters: \(s_{0}\sim 1\), \(\delta/2\pi=-3\) MHz from the Zeeman shifted cycling transition at \(B_{z}=20\) G, with polarization impurity of \(6.3\times 10^{-4}\), and equally distributed Zeeman state preparation. Results are shown in Fig. \ref{fig_transient}.

The scattering process is a sequence of  quantum jumps that can be modeled as a Markov chain with finite state-change probability based on the coupling strength \(|\Omega|^2\) and decay branching ratios.
The effects from time-dependence of the probe intensity and the ODTs are ignored in the simulation. We see a sharp increase in dark state probability until the light optically pumps the atoms. We note that the transient depumping, which gives an error in state determination, could be eliminated by coherently transferring the atoms from the qubit state $\ket{2,0}$ to the stretched state $\ket{2,2}$ using microwave or Raman pulses.

\section{Resonant and off-resonant scattering rates}

The Rabi frequency between a ground state $\Ket{n,l,J,F,m_F}$ and an excited state $\Ket{n',l',J',F',m_F'}$ is given by the expression
\begin{equation}
  \begin{split}
    \left|\Omega^{F',m_{F'}}_{F,m_{F}}\right|^{2} &=\left(\frac{e {\mathcal E}_{q}}{\hbar}\right)^{2}(2F+1)\\
    &\times \left|C^{F' m'_{F}}_{F m_{F} 1 q}
    \begin{Bmatrix}
      J & I & F\\
      F' & 1 & J'
    \end{Bmatrix}
    \Braket{n'l'J'||r||nlJ}\right|^{2}
  \end{split}
\end{equation}
where \(q=m_{F'}-m_{F}\), \(C^{F' m_{F'}}_{F m_{F} 1 q}\) is a Clebsch-Gordon coefficient, and \(\lbrace\rbrace\) is a Wigner 6j symbol.
 Spherical component $q$ of the optical field is $E_q=\frac{{\mathcal E}_q}{2}e^{-\imath \omega t}+ c.c.~$.
Only coupling to \(5p_{3/2}\) is considered due to the large fine structure splitting between $5p_{1/2}$ and $5p_{3/2}$. 
This allows the couplings to be expressed in terms of the common reduced matrix element \(\Braket{5p_{3/2}||r||5s_{1/2}}\).

Using the normal scattering rate equation for a two-level system, the scattering rate for each specific transition and polarization can be written as:
\begin{equation}
  r^{F',m_{F'}}_{F,m_{F}}=\frac{\gamma}{2}\frac{\frac{2|\Omega|^{2}}{\gamma^2}}{1+\frac{4\delta^{2}}{\gamma^{2}}+\frac{2|\Omega|^{2}}{\gamma^2}}\Bigr\rvert_{\Omega=\Omega^{F',m_{F'}}_{F,m_{F}}}.
\end{equation}
Atoms excited to \(\Ket{F'}\) levels can spontaneously decay to \(\Ket{F}\) with branching ratio given by:
\begin{equation}
  b^{F'}_{F}=(2J'+1)(2F+1)\left|
  \begin{Bmatrix}
  J & I & F\\
  F' & 1 & J'
  \end{Bmatrix}\right|^{2},
\end{equation}
which satisfies the normalization condition \(\sum\nolimits_{F}b^{F'}_{F}=1\).

After a single photon absorption-emission cycle of the bright state, the two possibilities are to decay back to  $\Ket{B}$ by a resonant process or to $\Ket{D}$ by an off-resonant Raman process.
In the following sections the relative strength between the cycling transition and the leakage into $\Ket{D}$ is calculated for the cases of unpolarized and circularly-polarized probe light.

\subsection{Unpolarized illumination}
For the case of probing Zeeman degenerate atoms with unpolarized probe light, the rates for resonant and off-resonant processes are obtained by summing over the scattering rate weighted by the branching ratios with initial ground level \(F_i=2\) and final levels \(F_f=2\) or 1.

\begin{equation}
  r_{i\rightarrow f}=\sum_{F',m_{F'}} \sum_{m_{F_{i}}} r^{F',m_{F'}}_{F_{i},m_{F_{i}}}b^{F'}_{F_{f}}
\end{equation}
We will denote the resonant process \(r_{c}=r_{2\rightarrow 2}\) and the off-resonant process as \(r_{R}=r_{2\rightarrow 1}\).
From the relative rate of both processes, we obtain a probability to depump for each scattered resonant photon \(\beta=r_{R}/r_{c}\).
Alternatively, we can express the rate in terms of the mean number of emitted photons before a depumping event as: \(N_{\gamma}=1/\beta\).
For near-resonant light \(\delta\ll\delta_{2'-3'}\) we obtain the approximate expression for \(N_{\gamma}\):
\begin{equation}
  N_{\gamma}=\frac{38340}{1+4\delta^2/\gamma^2+s_{0}},
\end{equation}
where $\delta$ is the probe detuning from the $2-3'$ transition, \(s_{0}=I/I_{s}\) and \(I_{s}=3.58 ~\textrm{mW}/\textrm{cm}^2\).

\subsection{Circularly-polarized illumination}

For the case of Zeeman non-degenerate atoms and circularly polarized probe light, the probe optically pumps atoms into the stretched state $\Ket{2,2}$ where most of the scattering events take place.
After the initial fast optical pumping, we can simplify the model to only consider transitions accessible from \(\ket{2,2}\).
The dominant transition will be the cycling transition \(\ket{2,2}\rightarrow\ket{3',3'}\) by design.
The other transitions are only accessible due to polarization and alignment imperfections, which open off-resonant transitions to \(\ket{2',2'}\), \(\ket{2',1'}\) and \(\ket{1',1'}\).
Since transitions to $\Ket{3',m_F'\neq 3'}$, will scatter only to $\Ket{2}$ and be pumped back to $\Ket{2,2}$, these states can be ignored in this analysis.

To see the influence and sensitivity of polarization and alignment errors, we parametrize the  polarization purity  by the  intensity fraction 
$(\epsilon_{-1}, \epsilon_{1})=(1-p,p)$ where $\epsilon_q$ denotes the fraction of spherical component $q$ and  $0\leq p \leq 1$. The angular mismatch between the quantization-axis $\hat{z}$ and $\vec{k}_{\rm RO}$ is represented by the angle $\theta$.
An angular mismatch \(\theta\) projects the circular polarization to the $z$ axis, allowing coupling to \(\Delta m_{F}=0\) transitions.
For a given polarization purity and angular mismatch, one can calculate the projection using the Wigner-D function for a spin-1 particle.
The perfectly matched case gives a projection \(\left(w_{-1},w_{0},w_{+1}\right)\)=\((1-p,0,p)\) where \(\sum_{q=-1}^1w_{q}=1\).
If \(\theta\) is small, the resonant process rate \(r_{c}\) can be written as
\begin{equation}
\begin{split}
r_{c}&=r^{3',3'}_{2,2}(w_{+1})\\
&\equiv \frac{\gamma}{2}\frac{s_{0}\left[p (D^{1}_{1,1}(\theta))^{2}+(1-p)(D^{1}_{-1,1}(\theta))^{2}\right]}{1+\frac{4\delta^{2}}{\gamma^{2}}+s_{0}\left[p (D^{1}_{1,1}(\theta))^{2}+(1-p)(D^{1}_{-1,1}(\theta))^{2}\right]}
\end{split}
\end{equation}
where \(D^{j}_{m_{1},m_{2}}(\theta)\) is the Wigner-D function and for the following derivation, they are explicitly written. Here \(s_{0}=I/I_{s,c}\), and \(I_{s,c}=1.66 ~\textrm{mW}/\textrm{cm}^2\).
Raman processes can occur via coupling to \(\ket{2',2'}\) by \(\pi\)-projection and to \(\ket{2',1'}\) and \(\ket{1',1'}\) by \(\sigma_{-}\) projection. Considering the branching ratios to the dark state, the Raman rate \(r_{R}\) is
\begin{equation}
r_{R}=\left[r^{2',2'}_{2,2}(w_{0})+r^{2',1'}_{2,2}(w_{-1})\right]b_{1}^{2'}+r^{1',1'}_{2,2}(w_{-1})b_{1}^{1'}
\end{equation}
Neglecting saturation effects on \(2\leftrightarrow 2'\) and \(2\leftrightarrow 1'\) transitions, we can simplify the denominators and obtain
\begin{equation}
\begin{split}
r_{R}&\simeq\frac{\gamma}{2}\frac{1}{2}\left[\frac{s_{0}}{3}\left(p\sin^{2}{\frac{\theta}{2}}+(1-p)\cos^{2}{\frac{\theta}{2}}\right)\right. \\
&\left. +\frac{s_{0}}{6}\left((1-p)\cos^{4}{\frac{\theta}{2}}+p\sin^{4}{\frac{\theta}{2}}\right)\right]\\
&/\left(1+\frac{4\delta_{2'-3'}^{2}}{\gamma^{2}}\right)\\
&+\frac{\gamma}{2}\frac{5}{6}\left[\frac{s_{0}}{6}\frac{(1-p)\cos^{4}{\theta/2}+p\sin^{4}{\theta/2}}{1+\frac{4\delta_{1'-3'}^{2}}{\gamma^{2}}}\right]
\end{split}
\end{equation}
where $\delta_{2'-3'}, \delta_{1'-3'}$ are excited state hyperfine splittings. The 
denominators can be further simplified at large detuning, yielding
\begin{equation}
r_{R}=\frac{\gamma s_{0}}{2} f(\theta,p)
\end{equation}
where
\begin{equation}
\begin{split}
f(\theta,p)&=\frac{\gamma^{2}}{48\delta^{2}_{2'-3'}}\left[2\left(p\sin^{2}{\frac{\theta}{2}}+(1-p)\cos^{2}{\frac{\theta}{2}}\right)\right.\\
&\left.+\left((1-p)\cos^{4}{\frac{\theta}{2}}+p\sin^{4}{\frac{\theta}{2}}\right)\right]\\
&+\frac{5\gamma^{2}}{144\delta^{2}_{1'-3'}}\left[(1-p)\cos^{4}{\theta/2}+p\sin^{4}{\theta/2}\right].
\end{split}
\end{equation}
Again, taking the relative rate \(r_{c}/r_{R}\) gives the mean number of photons the bright state can scatter before one depumping event.
This can be approximated by:
\begin{equation}
\begin{split}
N_{\gamma,\sigma}(\theta,p)&\simeq\frac{p\cos^{4}{\frac{\theta}{2}}}{p\sin^{4}{\frac{\theta}{2}}+\sin^{2}{\frac{\theta}{2}}+(1-p)\cos^{4}{\frac{\theta}{2}}}\\
&\times\left(\frac{7}{4}\right)\left[\frac{38340}{1+4\delta^2/\gamma^2+s_{0}}\right].
\end{split}
\end{equation}
valid when \(\theta\ll1\), \(\delta\ll\delta_{2'-3'}\), \(p\sim 1\). For the perfectly aligned case \(\theta=0\), depumping is suppressed by a factor of $\frac{4}{7}\frac{1-p}{p}$. Under non-zero magnetic field and optical potential one should include state dependent Zeeman and AC stark shifts in order to obtain more accurate results.

\section{Fictitious magnetic field}

Consider the AC stark shift \(\Delta E_{\psi}\) of an atom in state \(\ket{\psi}a\) illuminated by a single frequency optical field \(\omega\) written as
\begin{equation}
\Delta E_{\psi}=-\frac{1}{4}\alpha_{\psi}(\omega)\mathcal{E}^{2}
\end{equation}
where \(\alpha_{\psi}\) is the dynamic polarizability which can be decomposed into scalar(\(\alpha^{(0)}_{\psi}\)), vector(\(\alpha^{(1)}_{\psi}\)) and tensor(\(\alpha^{(2)}_{\psi}\)) contributions as
\begin{equation}
\begin{split}
\alpha_{nJF}(\omega)&=\alpha^{(0)}_{nJF}(\omega)+\mathcal{A}(\hat{k}\cdot\hat{z})m_{F}\alpha^{(1)}_{nJF}(\omega)\\
&+\left[\frac{3(\hat{p}\cdot\hat{z})^{2}-1}{2}\right]\frac{3m^{2}_{F}-F(F+1)}{F(2F-1)}\alpha^{(2)}_{nJF}(\omega).
\end{split}
\end{equation}
Here \(\mathcal{A}\) represents the circularity of light ranging continuously from \(1\)(Right handed) to \(-1\)(Left handed), and \(0\) for linear polarization.
Unit vectors \(\hat{k}, \hat{p}\) and \(\hat{z}\) denote the orientations of wave vector, electric field and quantization axis defined by the bias magnetic field.

The effect of a vector light shift is equivalent to having a static magnetic field \(B_{\rm fict}\). We obtain an equivalent field by equating the shift to Zeeman shifts
\begin{equation}
-\frac{1}{4}\mathcal{A}m_{F}\alpha^{(1)}_{nJF}\mathcal{E}^2\hat{k}=\mu_{B}g_{nJF}m_{F}\vec{B}_{\rm fict}
\end{equation}
the equivalent fictitious field is given by:
\begin{equation}
\vec{B}_{\rm fict}=-\frac{\mathcal{A}\alpha^{(1)}}{4\mu_{B}g_{nJF}}\mathcal{E}^{2}\hat{k}
\end{equation}
with Land\'{e} g-factor \(g_{nJF}\).

It is convenient to express the shift in terms of the mean trap depth for ground states as \(U_{\rm trap}=-1/4\alpha^{(0)}_{5s_{1/2}}\mathcal{E}^{2}\)
\begin{equation}
\vec{B}_{\rm fict}=\frac{U_{\rm trap}}{\mu_{B}g_{nJF}}
\left[\frac{\mathcal{A}\alpha^{(1)}_{5s_{1/2}}}{\alpha^{(0)}_{5s_{1/2}}}\right]\hat{k}
\end{equation}
The equation gives both magnitude and orientation of the fictitious magnetic field. With finite circularity of the  ODT light the bright state experiences
\begin{equation}
\frac{d\vec{B}_{\rm fict}}{dU_{\rm trap}}=29.77 \left[\frac{\mathcal{A}\alpha^{(1)}_{5s_{1/2}}}{\alpha^{(0)}_{5s_{1/2}}}\right]\hat{k}\quad \text{G}/\text{mK}
\end{equation}
where we have expressed the trap depth in temperature units.

The impact of this extra field is an effective time-dependent magnetic field during the 6 ms interrogation stage.
While atoms are being probed, the traps are off and they are continuously pumped and projected to the stretched state, \(\ket{2,2}\) along the quantization axis \( \hat{z}\) originally set by the external static magnetic field \(\vec{B}_{\rm ext}\).
When the traps are turned back on, \(\vec{B}_{\rm ODT}=\vec{B}_{\rm ext}+\vec{B}_{\rm fict}\) sets the quantization axis, and if these two axes are not parallel, \(\ket{2,2}\) is no longer an eigenstate.
The state $\Ket{2,2}$ will undergo Larmor precession about the new axis with the frequency \(\omega_{L}=\mu_{B}g_{F}{B}_{\rm ODT}/\hbar\).
The projection onto other Zeeman states in a rotated frame can be calculated from the Wigner-D function for a spin-2 particle.
If the atom is projected to a non-stretched state when the next probe cycle  begins, it will go through another optical pumping cycle and experience a temporarily increased depumping rate to \(\ket{D}\).

\begin{figure}[!t]
  \includegraphics[width=225pt]{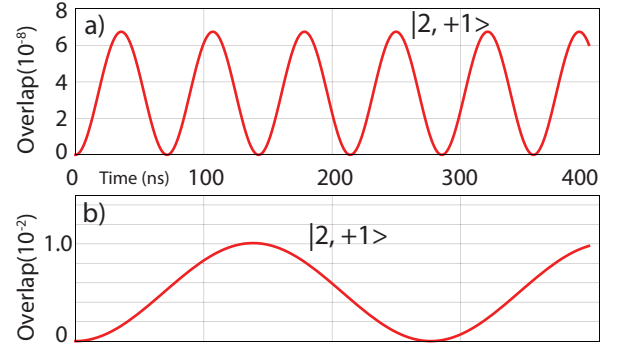}
  \caption{(color online)
   Larmor precession of \(\ket{2,+2}\) showing finite overlap to \(\ket{2,+1}\).
	a) For our experimental configuration: \(\alpha=60^{\circ}\),\(B_{ext}=20\textrm{G}\), \(B_{fict}=3 \textrm{mG}\).
    b) Experiment configuration with more fictitious field. \(B_{ext}=5 \textrm{G}\), \(B_{fict}=0.3 \textrm{G}\).
    Both cases have negligible overlap to \(m=-2, -1, 0\).
    }
   \label{fig_Larmor}
\end{figure}

To estimate the contribution of this effect to depumping let us begin with the initial stretched state \(\ket{2,2}\) and have it precess under magnetic field \(\vec{B}_{\rm ODT}\), governed by the Hamiltonian \(H=\mu_{B}g_{F}\vec{F}\cdot\vec{B}_{\textrm{ODT}}\) where $\vec{F}$ is the angular momentum operator. 
The angle \(\theta_{0}\) between \(\vec{B}_{\rm ext}\) and the new field \(\vec{B}_{\rm ODT}\) is
\begin{equation}
\theta_{0} = \arctan\left(\frac{x \sin\alpha}{1+x\cos\alpha}\right)
\label{vector_mismtch_angle}
\end{equation}
where \(\alpha\) is the angle between \(\vec{B}_{\rm ext}\) and \(\vec{B}_{\rm fict}\) and \(x\equiv B_{\rm fict}/B_{\rm ext}\) is the relative strength of the fields.
For small axis mismatch, precession couples the closest \(m\) state, \(\ket{2,2}\leftrightarrow\ket{2,1}\). Projection to the \(m_F=1\) state is time-dependent and dictated by the Larmor frequency. We consider maximal overlap to estimate maximal depumping due to the fictitious field.
\begin{equation}
\left|\bra{2,1}e^{\frac{-iHt}{\hbar}}\ket{2,2}\right|^2\leq \left|D^{2}_{2,1}(2\theta_0)\right|^2=4\cos^{6}{\theta_0}\sin^{2}{\theta_0}
\end{equation}
Substituting \(\theta_{0}\) obtained from Eq. (\ref{vector_mismtch_angle}) shows the overlap is bounded by
\begin{equation}
\left|\bra{2,1}e^{\frac{-iHt}{\hbar}}\ket{2,2}\right|^2\leq \frac{4x^{2}\left(1+x\cos^6\alpha\right)\sin^2{\alpha}}{\left(1+x^2+2x\cos\alpha\right)^{4}}.
\end{equation}
For small vector shift (\(x\ll 1\)), the expression reduces to \(\leq 4x^2\sin^2 (\alpha)\) showing a quadratic dependency.
For our experiment, the vector light shift makes an angle \(\alpha=60^{\circ}\), and the relative strength of the fields \(x\) is \(1.5\times 10^{-4}\).
These parameters give the maximal projection of precessed stretched state to the neighboring $m$-state to be \(6.7\times 10^{-8}\) per cycle, which agrees with the numerical simulation shown in Fig.\ref{fig_Larmor}a).
What this means is that when the precessed state is illuminated by the light again, it has a probability of being projected to the $m_F=1$ state of at most that number. Since we repeat the 800 ns long chopping cycle for 6 ms, \(\sim 7500\) projections will occur. Multiplying with 7500 gives an expected number of $m$-changing scattering events \(5.1\times 10^{-4}\) per readout, and each readout scatters \(\sim 10^{4}\) photons. Therefore the vector light shift induced depumping is equivalent to having polarization contamination at the  \(\sim 10^{-8}\) level, orders of magnitude smaller than our measured contamination. This agrees with our observation that we failed to observe trap intensity dependent depumping, once the circularity in the ODT light was reduced from the parameters in Fig \ref{fig_Larmor}b).


\begin{thebibliography}{23}%
\makeatletter
\providecommand \@ifxundefined [1]{%
 \@ifx{#1\undefined}
}%
\providecommand \@ifnum [1]{%
 \ifnum #1\expandafter \@firstoftwo
 \else \expandafter \@secondoftwo
 \fi
}%
\providecommand \@ifx [1]{%
 \ifx #1\expandafter \@firstoftwo
 \else \expandafter \@secondoftwo
 \fi
}%
\providecommand \natexlab [1]{#1}%
\providecommand \enquote  [1]{``#1''}%
\providecommand \bibnamefont  [1]{#1}%
\providecommand \bibfnamefont [1]{#1}%
\providecommand \citenamefont [1]{#1}%
\providecommand \href@noop [0]{\@secondoftwo}%
\providecommand \href [0]{\begingroup \@sanitize@url \@href}%
\providecommand \@href[1]{\@@startlink{#1}\@@href}%
\providecommand \@@href[1]{\endgroup#1\@@endlink}%
\providecommand \@sanitize@url [0]{\catcode `\\12\catcode `\$12\catcode
  `\&12\catcode `\#12\catcode `\^12\catcode `\_12\catcode `\%12\relax}%
\providecommand \@@startlink[1]{}%
\providecommand \@@endlink[0]{}%
\providecommand \url  [0]{\begingroup\@sanitize@url \@url }%
\providecommand \@url [1]{\endgroup\@href {#1}{\urlprefix }}%
\providecommand \urlprefix  [0]{URL }%
\providecommand \Eprint [0]{\href }%
\providecommand \doibase [0]{http://dx.doi.org/}%
\providecommand \selectlanguage [0]{\@gobble}%
\providecommand \bibinfo  [0]{\@secondoftwo}%
\providecommand \bibfield  [0]{\@secondoftwo}%
\providecommand \translation [1]{[#1]}%
\providecommand \BibitemOpen [0]{}%
\providecommand \bibitemStop [0]{}%
\providecommand \bibitemNoStop [0]{.\EOS\space}%
\providecommand \EOS [0]{\spacefactor3000\relax}%
\providecommand \BibitemShut  [1]{\csname bibitem#1\endcsname}%
\let\auto@bib@innerbib\@empty
\bibitem [{\citenamefont {Saffman}(2016)}]{Saffman2016}%
  \BibitemOpen
  \bibfield  {author} {\bibinfo {author} {\bibfnamefont {M.}~\bibnamefont
  {Saffman}},\ }\bibfield  {title} {\enquote {\bibinfo {title} {Quantum
  computing with atomic qubits and {R}ydberg interactions: Progress and
  challenges},}\ }\href@noop {} {\bibfield  {journal} {\bibinfo  {journal} {J.
  Phys. B}\ }\textbf {\bibinfo {volume} {49}},\ \bibinfo {pages} {202001}
  (\bibinfo {year} {2016})}\BibitemShut {NoStop}%
\bibitem [{\citenamefont {Endres}\ \emph {et~al.}(2016)\citenamefont {Endres},
  \citenamefont {Bernien}, \citenamefont {Keesling}, \citenamefont {Levine},
  \citenamefont {Anschuetz}, \citenamefont {Krajenbrink}, \citenamefont
  {Senko}, \citenamefont {Vuletic}, \citenamefont {Greiner},\ and\
  \citenamefont {Lukin}}]{Endres2016}%
  \BibitemOpen
  \bibfield  {author} {\bibinfo {author} {\bibfnamefont {M.}~\bibnamefont
  {Endres}}, \bibinfo {author} {\bibfnamefont {H.}~\bibnamefont {Bernien}},
  \bibinfo {author} {\bibfnamefont {A.}~\bibnamefont {Keesling}}, \bibinfo
  {author} {\bibfnamefont {H.}~\bibnamefont {Levine}}, \bibinfo {author}
  {\bibfnamefont {E.~R.}\ \bibnamefont {Anschuetz}}, \bibinfo {author}
  {\bibfnamefont {A.}~\bibnamefont {Krajenbrink}}, \bibinfo {author}
  {\bibfnamefont {C.}~\bibnamefont {Senko}}, \bibinfo {author} {\bibfnamefont
  {V.}~\bibnamefont {Vuletic}}, \bibinfo {author} {\bibfnamefont
  {M.}~\bibnamefont {Greiner}}, \ and\ \bibinfo {author} {\bibfnamefont
  {M.~D.}\ \bibnamefont {Lukin}},\ }\bibfield  {title} {\enquote {\bibinfo
  {title} {Atom-by-atom assembly of defect-free one-dimensional cold atom
  arrays},}\ }\href@noop {} {\bibfield  {journal} {\bibinfo  {journal}
  {Science}\ }\textbf {\bibinfo {volume} {354}},\ \bibinfo {pages} {1024}
  (\bibinfo {year} {2016})}\BibitemShut {NoStop}%
\bibitem [{\citenamefont {Xia}\ \emph {et~al.}(2015)\citenamefont {Xia},
  \citenamefont {Lichtman}, \citenamefont {Maller}, \citenamefont {Carr},
  \citenamefont {Piotrowicz}, \citenamefont {Isenhower},\ and\ \citenamefont
  {Saffman}}]{Xia2015}%
  \BibitemOpen
  \bibfield  {author} {\bibinfo {author} {\bibfnamefont {T.}~\bibnamefont
  {Xia}}, \bibinfo {author} {\bibfnamefont {M.}~\bibnamefont {Lichtman}},
  \bibinfo {author} {\bibfnamefont {K.}~\bibnamefont {Maller}}, \bibinfo
  {author} {\bibfnamefont {A.~W.}\ \bibnamefont {Carr}}, \bibinfo {author}
  {\bibfnamefont {M.~J.}\ \bibnamefont {Piotrowicz}}, \bibinfo {author}
  {\bibfnamefont {L.}~\bibnamefont {Isenhower}}, \ and\ \bibinfo {author}
  {\bibfnamefont {M.}~\bibnamefont {Saffman}},\ }\bibfield  {title} {\enquote
  {\bibinfo {title} {Randomized benchmarking of single-qubit gates in a {2D}
  array of neutral-atom qubits},}\ }\href@noop {} {\bibfield  {journal}
  {\bibinfo  {journal} {Phys. Rev. Lett.}\ }\textbf {\bibinfo {volume} {114}},\
  \bibinfo {pages} {100503} (\bibinfo {year} {2015})}\BibitemShut {NoStop}%
\bibitem [{\citenamefont {Barredo}\ \emph {et~al.}(2016)\citenamefont
  {Barredo}, \citenamefont {de~Les\'el\'euc}, \citenamefont {Lienhard},
  \citenamefont {Lahaye},\ and\ \citenamefont {Browaeys}}]{Barredo2016}%
  \BibitemOpen
  \bibfield  {author} {\bibinfo {author} {\bibfnamefont {D.}~\bibnamefont
  {Barredo}}, \bibinfo {author} {\bibfnamefont {S.}~\bibnamefont
  {de~Les\'el\'euc}}, \bibinfo {author} {\bibfnamefont {V.}~\bibnamefont
  {Lienhard}}, \bibinfo {author} {\bibfnamefont {T.}~\bibnamefont {Lahaye}}, \
  and\ \bibinfo {author} {\bibfnamefont {A.}~\bibnamefont {Browaeys}},\
  }\bibfield  {title} {\enquote {\bibinfo {title} {An atom-by-atom assembler of
  defect-free arbitrary two-dimensional atomic arrays},}\ }\href@noop {}
  {\bibfield  {journal} {\bibinfo  {journal} {Science}\ }\textbf {\bibinfo
  {volume} {354}},\ \bibinfo {pages} {1021} (\bibinfo {year}
  {2016})}\BibitemShut {NoStop}%
\bibitem [{\citenamefont {Wang}\ \emph {et~al.}(2016)\citenamefont {Wang},
  \citenamefont {Kumar}, \citenamefont {Wu},\ and\ \citenamefont
  {Weiss}}]{YWang2016}%
  \BibitemOpen
  \bibfield  {author} {\bibinfo {author} {\bibfnamefont {Y.}~\bibnamefont
  {Wang}}, \bibinfo {author} {\bibfnamefont {A.}~\bibnamefont {Kumar}},
  \bibinfo {author} {\bibfnamefont {T.-Y.}\ \bibnamefont {Wu}}, \ and\ \bibinfo
  {author} {\bibfnamefont {D.~S.}\ \bibnamefont {Weiss}},\ }\bibfield  {title}
  {\enquote {\bibinfo {title} {Single-qubit gates based on targeted phase
  shifts in a 3{D} neutral atom array},}\ }\href@noop {} {\bibfield  {journal}
  {\bibinfo  {journal} {Science}\ }\textbf {\bibinfo {volume} {352}},\ \bibinfo
  {pages} {1562} (\bibinfo {year} {2016})}\BibitemShut {NoStop}%
\bibitem [{\citenamefont {Devitt}\ \emph {et~al.}(2013)\citenamefont {Devitt},
  \citenamefont {Munro},\ and\ \citenamefont {Nemoto}}]{Devitt2013}%
  \BibitemOpen
  \bibfield  {author} {\bibinfo {author} {\bibfnamefont {S.~J.}\ \bibnamefont
  {Devitt}}, \bibinfo {author} {\bibfnamefont {W.~J.}\ \bibnamefont {Munro}}, \
  and\ \bibinfo {author} {\bibfnamefont {K.}~\bibnamefont {Nemoto}},\
  }\bibfield  {title} {\enquote {\bibinfo {title} {Quantum error correction for
  beginners},}\ }\href@noop {} {\bibfield  {journal} {\bibinfo  {journal} {Rep.
  Prog. Phys.}\ }\textbf {\bibinfo {volume} {76}},\ \bibinfo {pages} {076001}
  (\bibinfo {year} {2013})}\BibitemShut {NoStop}%
\bibitem [{\citenamefont {Myerson}\ \emph {et~al.}(2008)\citenamefont
  {Myerson}, \citenamefont {Szwer}, \citenamefont {Webster}, \citenamefont
  {Allcock}, \citenamefont {Curtis}, \citenamefont {Imreh}, \citenamefont
  {Sherman}, \citenamefont {Stacey}, \citenamefont {Steane},\ and\
  \citenamefont {Lucas}}]{Myerson2008}%
  \BibitemOpen
  \bibfield  {author} {\bibinfo {author} {\bibfnamefont {A.~H.}\ \bibnamefont
  {Myerson}}, \bibinfo {author} {\bibfnamefont {D.~J.}\ \bibnamefont {Szwer}},
  \bibinfo {author} {\bibfnamefont {S.~C.}\ \bibnamefont {Webster}}, \bibinfo
  {author} {\bibfnamefont {D.~T.~C.}\ \bibnamefont {Allcock}}, \bibinfo
  {author} {\bibfnamefont {M.~J.}\ \bibnamefont {Curtis}}, \bibinfo {author}
  {\bibfnamefont {G.}~\bibnamefont {Imreh}}, \bibinfo {author} {\bibfnamefont
  {J.~A.}\ \bibnamefont {Sherman}}, \bibinfo {author} {\bibfnamefont {D.~N.}\
  \bibnamefont {Stacey}}, \bibinfo {author} {\bibfnamefont {A.~M.}\
  \bibnamefont {Steane}}, \ and\ \bibinfo {author} {\bibfnamefont {D.~M.}\
  \bibnamefont {Lucas}},\ }\bibfield  {title} {\enquote {\bibinfo {title}
  {High-fidelity readout of trapped-ion qubits},}\ }\href@noop {} {\bibfield
  {journal} {\bibinfo  {journal} {Phys. Rev. Lett.}\ }\textbf {\bibinfo
  {volume} {100}},\ \bibinfo {pages} {200502} (\bibinfo {year}
  {2008})}\BibitemShut {NoStop}%
\bibitem [{\citenamefont {H\"affner}\ \emph {et~al.}(2008)\citenamefont
  {H\"affner}, \citenamefont {Roos},\ and\ \citenamefont
  {Blatt}}]{Haffner2008}%
  \BibitemOpen
  \bibfield  {author} {\bibinfo {author} {\bibfnamefont {H.}~\bibnamefont
  {H\"affner}}, \bibinfo {author} {\bibfnamefont {C.~F.}\ \bibnamefont {Roos}},
  \ and\ \bibinfo {author} {\bibfnamefont {R.}~\bibnamefont {Blatt}},\
  }\bibfield  {title} {\enquote {\bibinfo {title} {Quantum computing with
  trapped ions},}\ }\href@noop {} {\bibfield  {journal} {\bibinfo  {journal}
  {Phys. Rep.}\ }\textbf {\bibinfo {volume} {469}},\ \bibinfo {pages} {155}
  (\bibinfo {year} {2008})}\BibitemShut {NoStop}%
\bibitem [{\citenamefont {Fuhrmanek}\ \emph {et~al.}(2011)\citenamefont
  {Fuhrmanek}, \citenamefont {Bourgain}, \citenamefont {Sortais},\ and\
  \citenamefont {Browaeys}}]{Fuhrmanek2011}%
  \BibitemOpen
  \bibfield  {author} {\bibinfo {author} {\bibfnamefont {A.}~\bibnamefont
  {Fuhrmanek}}, \bibinfo {author} {\bibfnamefont {R.}~\bibnamefont {Bourgain}},
  \bibinfo {author} {\bibfnamefont {Y.~R.~P.}\ \bibnamefont {Sortais}}, \ and\
  \bibinfo {author} {\bibfnamefont {A.}~\bibnamefont {Browaeys}},\ }\bibfield
  {title} {\enquote {\bibinfo {title} {Free-space lossless state detection of a
  single trapped atom},}\ }\href@noop {} {\bibfield  {journal} {\bibinfo
  {journal} {Phys. Rev. Lett.}\ }\textbf {\bibinfo {volume} {106}},\ \bibinfo
  {pages} {133003} (\bibinfo {year} {2011})}\BibitemShut {NoStop}%
\bibitem [{\citenamefont {Gibbons}\ \emph {et~al.}(2011)\citenamefont
  {Gibbons}, \citenamefont {Hamley}, \citenamefont {Shih},\ and\ \citenamefont
  {Chapman}}]{Gibbons2011}%
  \BibitemOpen
  \bibfield  {author} {\bibinfo {author} {\bibfnamefont {M.~J.}\ \bibnamefont
  {Gibbons}}, \bibinfo {author} {\bibfnamefont {C.~D.}\ \bibnamefont {Hamley}},
  \bibinfo {author} {\bibfnamefont {C.-Y.}\ \bibnamefont {Shih}}, \ and\
  \bibinfo {author} {\bibfnamefont {M.~S.}\ \bibnamefont {Chapman}},\
  }\bibfield  {title} {\enquote {\bibinfo {title} {Nondestructive fluorescent
  state detection of single neutral atom qubits},}\ }\href@noop {} {\bibfield
  {journal} {\bibinfo  {journal} {Phys. Rev. Lett.}\ }\textbf {\bibinfo
  {volume} {106}},\ \bibinfo {pages} {133002} (\bibinfo {year}
  {2011})}\BibitemShut {NoStop}%
\bibitem [{\citenamefont {Jau}\ \emph {et~al.}(2016)\citenamefont {Jau},
  \citenamefont {Hankin}, \citenamefont {Keating}, \citenamefont {Deutsch},\
  and\ \citenamefont {Biedermann}}]{Jau2016}%
  \BibitemOpen
  \bibfield  {author} {\bibinfo {author} {\bibfnamefont {Y.-Y.}\ \bibnamefont
  {Jau}}, \bibinfo {author} {\bibfnamefont {A.~M.}\ \bibnamefont {Hankin}},
  \bibinfo {author} {\bibfnamefont {T.}~\bibnamefont {Keating}}, \bibinfo
  {author} {\bibfnamefont {I.~H.}\ \bibnamefont {Deutsch}}, \ and\ \bibinfo
  {author} {\bibfnamefont {G.~W.}\ \bibnamefont {Biedermann}},\ }\bibfield
  {title} {\enquote {\bibinfo {title} {Entangling atomic spins with a
  {R}ydberg-dressed spin-flip blockade},}\ }\href@noop {} {\bibfield  {journal}
  {\bibinfo  {journal} {Nat. Phys.}\ }\textbf {\bibinfo {volume} {12}},\
  \bibinfo {pages} {71} (\bibinfo {year} {2016})}\BibitemShut {NoStop}%
\bibitem [{\citenamefont {Bochmann}\ \emph {et~al.}(2010)\citenamefont
  {Bochmann}, \citenamefont {M\"ucke}, \citenamefont {Guhl}, \citenamefont
  {Ritter}, \citenamefont {Rempe},\ and\ \citenamefont
  {Moehring}}]{Bochmann2010}%
  \BibitemOpen
  \bibfield  {author} {\bibinfo {author} {\bibfnamefont {J.}~\bibnamefont
  {Bochmann}}, \bibinfo {author} {\bibfnamefont {M.}~\bibnamefont {M\"ucke}},
  \bibinfo {author} {\bibfnamefont {C.}~\bibnamefont {Guhl}}, \bibinfo {author}
  {\bibfnamefont {S.}~\bibnamefont {Ritter}}, \bibinfo {author} {\bibfnamefont
  {G.}~\bibnamefont {Rempe}}, \ and\ \bibinfo {author} {\bibfnamefont {D.~L.}\
  \bibnamefont {Moehring}},\ }\bibfield  {title} {\enquote {\bibinfo {title}
  {Lossless state detection of single neutral atoms},}\ }\href@noop {}
  {\bibfield  {journal} {\bibinfo  {journal} {Phys. Rev. Lett.}\ }\textbf
  {\bibinfo {volume} {104}},\ \bibinfo {pages} {203601} (\bibinfo {year}
  {2010})}\BibitemShut {NoStop}%
\bibitem [{\citenamefont {Volz}\ \emph {et~al.}(2011)\citenamefont {Volz},
  \citenamefont {Gehr}, \citenamefont {Dubois}, \citenamefont {Est\`eve},\ and\
  \citenamefont {Reichel}}]{Volz2011}%
  \BibitemOpen
  \bibfield  {author} {\bibinfo {author} {\bibfnamefont {J.}~\bibnamefont
  {Volz}}, \bibinfo {author} {\bibfnamefont {R.}~\bibnamefont {Gehr}}, \bibinfo
  {author} {\bibfnamefont {G.}~\bibnamefont {Dubois}}, \bibinfo {author}
  {\bibfnamefont {J.}~\bibnamefont {Est\`eve}}, \ and\ \bibinfo {author}
  {\bibfnamefont {J.}~\bibnamefont {Reichel}},\ }\bibfield  {title} {\enquote
  {\bibinfo {title} {Measurement of the internal state of a single atom without
  energy exchange},}\ }\href@noop {} {\bibfield  {journal} {\bibinfo  {journal}
  {Nature}\ }\textbf {\bibinfo {volume} {475}},\ \bibinfo {pages} {210}
  (\bibinfo {year} {2011})}\BibitemShut {NoStop}%
\bibitem [{\citenamefont {Zhang}\ \emph {et~al.}(2012)\citenamefont {Zhang},
  \citenamefont {McConnell}, \citenamefont {\ifmmode~\acute{C}\else
  \'{C}\fi{}uk}, \citenamefont {Lin}, \citenamefont {Schleier-Smith},
  \citenamefont {Leroux},\ and\ \citenamefont {Vuleti\ifmmode~\acute{c}\else
  \'{c}\fi{}}}]{HZhang2012}%
  \BibitemOpen
  \bibfield  {author} {\bibinfo {author} {\bibfnamefont {H.}~\bibnamefont
  {Zhang}}, \bibinfo {author} {\bibfnamefont {R.}~\bibnamefont {McConnell}},
  \bibinfo {author} {\bibfnamefont {S.}~\bibnamefont {\ifmmode~\acute{C}\else
  \'{C}\fi{}uk}}, \bibinfo {author} {\bibfnamefont {Q.}~\bibnamefont {Lin}},
  \bibinfo {author} {\bibfnamefont {M.~H.}\ \bibnamefont {Schleier-Smith}},
  \bibinfo {author} {\bibfnamefont {I.~D.}\ \bibnamefont {Leroux}}, \ and\
  \bibinfo {author} {\bibfnamefont {V.}~\bibnamefont
  {Vuleti\ifmmode~\acute{c}\else \'{c}\fi{}}},\ }\bibfield  {title} {\enquote
  {\bibinfo {title} {Collective state measurement of mesoscopic ensembles with
  single-atom resolution},}\ }\href@noop {} {\bibfield  {journal} {\bibinfo
  {journal} {Phys. Rev. Lett.}\ }\textbf {\bibinfo {volume} {109}},\ \bibinfo
  {pages} {133603} (\bibinfo {year} {2012})}\BibitemShut {NoStop}%
\bibitem [{\citenamefont {Saffman}\ and\ \citenamefont
  {Walker}(2005)}]{Saffman2005b}%
  \BibitemOpen
  \bibfield  {author} {\bibinfo {author} {\bibfnamefont {M.}~\bibnamefont
  {Saffman}}\ and\ \bibinfo {author} {\bibfnamefont {T.~G.}\ \bibnamefont
  {Walker}},\ }\bibfield  {title} {\enquote {\bibinfo {title} {Entangling
  single- and {N}-atom qubits for fast quantum state detection and
  transmission},}\ }\href@noop {} {\bibfield  {journal} {\bibinfo  {journal}
  {Phys. Rev. A}\ }\textbf {\bibinfo {volume} {72}},\ \bibinfo {pages} {042302}
  (\bibinfo {year} {2005})}\BibitemShut {NoStop}%
\bibitem [{\citenamefont {Beterov}\ and\ \citenamefont
  {Saffman}(2015)}]{Beterov2015}%
  \BibitemOpen
  \bibfield  {author} {\bibinfo {author} {\bibfnamefont {I.~I.}\ \bibnamefont
  {Beterov}}\ and\ \bibinfo {author} {\bibfnamefont {M.}~\bibnamefont
  {Saffman}},\ }\bibfield  {title} {\enquote {\bibinfo {title} {{R}ydberg
  blockade, {F}\"orster resonances, and quantum state measurements with
  different atomic species},}\ }\href@noop {} {\bibfield  {journal} {\bibinfo
  {journal} {Phys. Rev. A}\ }\textbf {\bibinfo {volume} {92}},\ \bibinfo
  {pages} {042710} (\bibinfo {year} {2015})}\BibitemShut {NoStop}%
\bibitem [{\citenamefont {Martinez-Dorantes}\ \emph {et~al.}(2017)\citenamefont
  {Martinez-Dorantes}, \citenamefont {W.Alt}, \citenamefont {Gallego},
  \citenamefont {Ghosh}, \citenamefont {Ratschbacher}, \citenamefont
  {V\"olzke},\ and\ \citenamefont {Meschede}}]{Martinez-Dorantes2017}%
  \BibitemOpen
  \bibfield  {author} {\bibinfo {author} {\bibfnamefont {M.}~\bibnamefont
  {Martinez-Dorantes}}, \bibinfo {author} {\bibnamefont {W.Alt}}, \bibinfo
  {author} {\bibfnamefont {J.}~\bibnamefont {Gallego}}, \bibinfo {author}
  {\bibfnamefont {S.}~\bibnamefont {Ghosh}}, \bibinfo {author} {\bibfnamefont
  {L.}~\bibnamefont {Ratschbacher}}, \bibinfo {author} {\bibfnamefont
  {Y.}~\bibnamefont {V\"olzke}}, \ and\ \bibinfo {author} {\bibfnamefont
  {D.}~\bibnamefont {Meschede}},\ }\bibfield  {title} {\enquote {\bibinfo
  {title} {Non-destructive parallel readout of neutral atom registers in
  optical potentials},}\ }\href@noop {} {\bibfield  {journal} {\bibinfo
  {journal} {arXiv:1706.00264}\ } (\bibinfo {year} {2017})}\BibitemShut
  {NoStop}%
\bibitem [{Kwo()}]{Kwon2017SM}%
  \BibitemOpen
  \href@noop {} {}\bibinfo {note} {See Supplemental Material at ??? for
  analysis of photo-electron statistics, camera noise, and atomic fluorescence
  dynamics with dependencies on experimental parameters.}\BibitemShut {Stop}%
\bibitem [{\citenamefont {Wineland}\ and\ \citenamefont
  {Itano}(1979)}]{Wineland1979}%
  \BibitemOpen
  \bibfield  {author} {\bibinfo {author} {\bibfnamefont {D.~J.}\ \bibnamefont
  {Wineland}}\ and\ \bibinfo {author} {\bibfnamefont {W.~M.}\ \bibnamefont
  {Itano}},\ }\bibfield  {title} {\enquote {\bibinfo {title} {Laser cooling of
  atoms},}\ }\href@noop {} {\bibfield  {journal} {\bibinfo  {journal} {Phys.
  Rev. A}\ }\textbf {\bibinfo {volume} {20}},\ \bibinfo {pages} {1521--1540}
  (\bibinfo {year} {1979})}\BibitemShut {NoStop}%
\bibitem [{\citenamefont {Piotrowicz}\ \emph {et~al.}(2013)\citenamefont
  {Piotrowicz}, \citenamefont {Lichtman}, \citenamefont {Maller}, \citenamefont
  {Li}, \citenamefont {Zhang}, \citenamefont {Isenhower},\ and\ \citenamefont
  {Saffman}}]{Piotrowicz2013}%
  \BibitemOpen
  \bibfield  {author} {\bibinfo {author} {\bibfnamefont {M.~J.}\ \bibnamefont
  {Piotrowicz}}, \bibinfo {author} {\bibfnamefont {M.}~\bibnamefont
  {Lichtman}}, \bibinfo {author} {\bibfnamefont {K.}~\bibnamefont {Maller}},
  \bibinfo {author} {\bibfnamefont {G.}~\bibnamefont {Li}}, \bibinfo {author}
  {\bibfnamefont {S.}~\bibnamefont {Zhang}}, \bibinfo {author} {\bibfnamefont
  {L.}~\bibnamefont {Isenhower}}, \ and\ \bibinfo {author} {\bibfnamefont
  {M.}~\bibnamefont {Saffman}},\ }\bibfield  {title} {\enquote {\bibinfo
  {title} {Two-dimensional lattice of blue-detuned atom traps using a projected
  {G}aussian beam array},}\ }\href@noop {} {\bibfield  {journal} {\bibinfo
  {journal} {Phys. Rev. A}\ }\textbf {\bibinfo {volume} {88}},\ \bibinfo
  {pages} {013420} (\bibinfo {year} {2013})}\BibitemShut {NoStop}%
\bibitem [{\citenamefont {Hirsch}\ \emph {et~al.}(2013)\citenamefont {Hirsch},
  \citenamefont {Wareham}, \citenamefont {Martin-Fernandez}, \citenamefont
  {Hobson},\ and\ \citenamefont {Rolfe}}]{Hirsch2013}%
  \BibitemOpen
  \bibfield  {author} {\bibinfo {author} {\bibfnamefont {M.}~\bibnamefont
  {Hirsch}}, \bibinfo {author} {\bibfnamefont {R.~J.}\ \bibnamefont {Wareham}},
  \bibinfo {author} {\bibfnamefont {M.~L.}\ \bibnamefont {Martin-Fernandez}},
  \bibinfo {author} {\bibfnamefont {M.~P.}\ \bibnamefont {Hobson}}, \ and\
  \bibinfo {author} {\bibfnamefont {D.~J.}\ \bibnamefont {Rolfe}},\ }\bibfield
  {title} {\enquote {\bibinfo {title} {A stochastic model for electron
  multiplication charge-coupled devices – from theory to practice},}\
  }\href@noop {} {\bibfield  {journal} {\bibinfo  {journal} {PLOS ONE}\
  }\textbf {\bibinfo {volume} {8}},\ \bibinfo {pages} {1--13} (\bibinfo {year}
  {2013})}\BibitemShut {NoStop}%
\bibitem [{\citenamefont {Hyv\"arinen}\ and\ \citenamefont
  {Oja}(2000)}]{Hyvarinen2000}%
  \BibitemOpen
  \bibfield  {author} {\bibinfo {author} {\bibfnamefont {A.}~\bibnamefont
  {Hyv\"arinen}}\ and\ \bibinfo {author} {\bibfnamefont {E.}~\bibnamefont
  {Oja}},\ }\bibfield  {title} {\enquote {\bibinfo {title} {Independent
  component analysis: algorithms and applications},}\ }\href@noop {} {\bibfield
   {journal} {\bibinfo  {journal} {Neural Netw.}\ }\textbf {\bibinfo {volume}
  {13}},\ \bibinfo {pages} {411 -- 430} (\bibinfo {year} {2000})}\BibitemShut
  {NoStop}%
\bibitem [{\citenamefont {Mart\'inez-Dorantes}(2016)}]{Martinez2016thesis}%
  \BibitemOpen
  \bibfield  {author} {\bibinfo {author} {\bibfnamefont {M.}~\bibnamefont
  {Mart\'inez-Dorantes}},\ }\emph {\bibinfo {title} {Fast non-destructive
  internal state detection of neutral atoms in optical potentials}},\
  \href@noop {} {Ph.D. thesis},\ \bibinfo  {school} {Rheinischen
  Friedrich-Wilhelms-Universit\"at Bonn} (\bibinfo {year} {2016})\BibitemShut
  {NoStop}%
\end{thebibliography}
\end{document}